\documentclass[aps,prl,twocolumn,amsmath,showpacs]{revtex4}

\def\Tr{\mbox{Tr}}

\begin{document}

\title{A unique quasi-probability for projective yes-no measurements}
\author{Lars M. Johansen}

\affiliation{Department of Technology, Buskerud University College,
N-3601 Kongsberg, Norway} \email{lars.m.johansen@hibu.no}
\date{\today}

\begin{abstract}
From an analysis of projective measurements, it is shown that the Wigner rule is the unique operational quasi-probability for the post-measurement state. A unique pre-measurement quasi-probability is derived from a principle of invariance of measurement disturbance under orthogonal projector complementation. Physical arguments for this principle are given. The informationally complete complex extension of the quasi-probability is also derived. Nonclassicality of this quasi-probability is due to measurement disturbance. The same quasi-probability follows from weak measurements.
\end{abstract}



\maketitle

In its most general form, classical physics deals with probability distributions over a Boolean logic of events. Quantum mechanics is also a probabilistic theory.  Every statement the theory makes about observations is probabilistic. However, it is not a classical probabilistic theory \cite{Bell-EinsPodoRosepara:64,Wigner-QuanMechDistFunc:71}. Successive measurements give probabilities dependent on measurement order. This is due to measurement disturbance.  Measurement disturbance in itself is not a nonclassical phenomenon. But in quantum mechanics, measurement disturbance is of an entirely different nature than in classical physics. It is not due to an imperfection of the instrument, but due to the invasive nature of quantum measurements. Classically, if one has sufficient information about how a measurement disturbs a system, it may be possible to compensate or subtract this disturbance.  But is compensation of measurement disturbance possible also in quantum mechanics? Obviously, it requires a  closer characterization of the quantum mechanical measurement disturbance. However, the prospects of such compensation or subtraction might seem dim \cite{Schwinger-QuanMech:01a}.

Wigner found a classical-like representation of quantum mechanics in terms of a quasi-probability over phase space \cite{Wigner-QuanCorrTherEqui:32}. The essential nonclassical aspect of the Wigner distribution is that it may take negative values. The Wigner distribution, together with a number of other quasi-probability distributions, have become useful tools for distinguishing quantum and classical effects. In finite dimensional Hilbert space, generalizations of the Wigner distribution have also been proposed (see \cite{Gibbons+Wootters-Discphasspacbase:04} and references therein). The Wigner distribution may be derived from a set of classical conditions. A basic condition is that unitary time evolution should correspond to classical propagation in phase space  \cite{Wigner-QuanMechDistFunc:71}. Whereas unitary evolution is associated with a closed system that is not subject to observation, measurements lead to non-unitary evolution. In this Letter, we derive a unique extension of the classical joint probability concept by analyzing projective measurements.  We shall proceed by imposing conditions motivated by physical arguments. This leads, in the end, to a complex quasi-probability that was originally discussed by Dirac \cite{Dirac-AnalBetwClasQuan:45}.

If negative ``probabilities'' is what it takes to make quantum mechanics look like a classical theory \cite{Feynman-SimuPhyswithComp:82}, it would be of interest to find an operational explanation of such a phenomenon. The Wigner distribution itself does not seem to give an answer. Although the Wigner distribution may be reconstructed using a variety of methods \cite{Leonhardt-MeasQuanStatLigh:97,Vogel+WelschETAL-QuanOpti:01}, a direct physical interpretation of negativity of the Wigner distribution in terms of measurements of the phase space observables themselves has not been obtained. With the quasi-probability derived in this Letter, we find a direct physical interpretation of nonclassical (i.e. negative and imaginary) quasi-probabilities in terms of measurement disturbance.

The discovery of weak values due to weak measurements \cite{Aharonov+AlbertETAL-ResuMeasCompSpin:88} came as a surprise to many. The notion that a measurement somehow could give values outside the eigenvalue spectrum seemed heretic. However, within the quasi-probability formalism derived here, both projective and weak measurements appear as consistent manifestations of the same underlying ``reality''. In fact, in the formalism derived here weak values appear as a natural consequence of projective measurements.

We will consider projective measurements \cite{Lueders-UEbeZust:51} in a Hilbert space of arbitrary, finite dimensions. Projective measurements are ideal, in the sense that they are repeatable while disturbing the system as little as possible. Any ideal measurement, quantum or classical, may be decomposed into elementary yes-no measurements. In quantum mechanics, such measurements are represented by projectors, the quantum generalization of classical events. A projector $\alpha^2=\alpha$ has eigenvalues 1 and 0, corresponding to outcomes yes or no, true or false. The probability of a yes in the projective measurement of this projector on a system prepared in a state $\rho$ is $\Tr \rho \alpha$.  The joint probability for positive outcomes in successive projective measurements of the two projectors $\alpha$ and $\beta$ is given by the Wigner rule $\Tr \rho \alpha \beta \alpha$ \cite{Wigner-ProbMeas:63}. A measurement in the opposite order gives the probability $\Tr \rho \beta \alpha \beta$. If the two projectors commute, both of these joint probabilities reduce to the order independent joint probability $\Tr \rho \alpha \beta$. If the two projectors do not commute, the expression $\Tr \rho \alpha \beta$ is in general complex. This does not make much sense as a joint probability \cite{Dirac-AnalBetwClasQuan:45}. Or does it?

We seek an operational quasi-probability ${\cal F}(\rho,\alpha,\beta)$. Operational here means that it should be connected as closely as possible to the classical definition of a joint probability in terms of successive measurements. The order of the arguments refer to an initial preparation $\rho$, a first measurement $\alpha$ and a second measurement $\beta$. We also introduce the orthogonal complements $\tilde{\alpha} = 1-\alpha$ and $\tilde{\beta}=1-\beta$. In correspondence with classical probability theory, the first condition on the quasi-probability is:

\paragraph{Condition 1.}
\begin{subequations}
\begin{eqnarray}
      {\cal F}(\rho,\alpha,\beta) + {\cal F}(\rho,\alpha,\tilde{\beta}) &=& \Tr \rho \alpha, \\
      {\cal F}(\rho,\alpha,\beta) + {\cal F}(\rho,\tilde{\alpha},\beta) &=& \Tr \rho \beta.
      \label{eq:correctmarginals2}
\end{eqnarray}
    \label{eq:correctmarginals}
\end{subequations}
This, of course, implies that the distribution is normalized to unity.

After a projective measurement of a projector $\alpha$, the ensemble corresponding to the eigenvalue 1 is represented by the state $\alpha \rho \alpha/\Tr \rho \alpha$  \cite{Lueders-UEbeZust:51}. The complete initial ensemble is represented after the measurement by the state \cite{Lueders-UEbeZust:51}
\begin{equation}
    \Lambda_\alpha(\rho) = \alpha \rho \alpha + \tilde{\alpha} \rho \tilde{\alpha}.
    \label{eq:lueders}
\end{equation}
We shall refer to such measurements as nonselective, and we shall refer to the state transformation $\Lambda_\alpha$ as the L\"uders map.

We first find the post-measurement quasi-probability ${\cal F} [\Lambda_\alpha(\rho),\alpha,\beta]$. A state $\sigma$ is undisturbed by a measurement $\Lambda_\alpha$ if $\Lambda_\alpha(\sigma)=\sigma$. The state $\Lambda_\alpha(\rho)$ is undisturbed by a measurement $\Lambda_\alpha$ since $\Lambda_\alpha[\Lambda_\alpha(\rho)]=\Lambda_\alpha(\rho)$. A subsequent measurement of $\beta$ can be made without it being disturbed by the preceding $\alpha$ measurement. The joint probability obtained in a successive measurement of $\alpha$ and $\beta$ on $\Lambda_\alpha(\rho)$ will not be influenced by measurement disturbance. We may therefore identify this joint probability with ${\cal F} [\Lambda_\alpha(\rho),\alpha,\beta]$. The Wigner rule applied to the state $\Lambda_\alpha(\rho)$ is ${\cal F} [\Lambda_\alpha(\rho),\alpha,\beta] = \Tr \Lambda_\alpha(\rho) \alpha \beta \alpha$. This simplifies to:
\paragraph{Condition 2.}
\begin{equation}
    {\cal F} [\Lambda_\alpha(\rho),\alpha,\beta] = \Tr \rho \alpha \beta \alpha.
    \label{eq:postdistribution}
\end{equation}
It is easily verified that this quasi-probability satisfies the marginality conditions (\ref{eq:correctmarginals}). This is due to the lack of measurement disturbance. Of course, Eq. (\ref{eq:postdistribution}) refers to a particular measurement order. As such, this is not a joint probability in the full classical sense. It is reasonable to assume that a \emph{pre-measurement} quasi-probability should not depend on the order in which subsequent measurements are performed, so that
\begin{equation}
    {\cal F}(\rho,\alpha,\beta) = {\cal F}(\rho,\beta,\alpha).
    \label{eq:ordersymmetry}
\end{equation}
We may note that the distribution  $\Tr \rho \alpha \Tr \rho \beta$ satisfies both this symmetry relation and the marginality conditions (\ref{eq:correctmarginals}). As such it is  a joint probability which is even nonnegative. However, it does not fulfill the condition (\ref{eq:postdistribution}).

We shall not impose the order symmetry (\ref{eq:ordersymmetry}) in the derivation of the quasi-probability ${\cal F}(\rho,\alpha,\beta)$. However, we shall use it to argue for another symmetry condition. To this end, we introduce the ``disturbance'' or the change of the quasi-probability ${\cal F}$ due to the L\"uders map $\Lambda_\alpha$,
\begin{equation}
    \Delta {\cal F} (\rho,\alpha,\beta) = {\cal F}(\rho,\alpha,\beta) -  {\cal F} [\Lambda_\alpha(\rho),\alpha,\beta].
    \label{eq:realquasi}
\end{equation}
We apply the symmetry condition (\ref{eq:ordersymmetry}) to the state $\Lambda_\alpha(\rho)$,
\begin{equation}
    {\cal F} [\Lambda_\alpha(\rho),\alpha,\beta] = {\cal F} [\Lambda_\alpha(\rho),\beta,\alpha].
\end{equation}
By using Eqs. (\ref{eq:postdistribution}) and (\ref{eq:realquasi}) this implies that
\begin{equation}
    \Tr \rho \alpha \beta \alpha = \Tr \Lambda_\alpha(\rho) \beta \alpha \beta + \Delta {\cal F} [ \Lambda_\alpha(\rho), \beta,\alpha].
    \label{eq:case1}
\end{equation}
Likewise, by exchanging $\beta$ with its orthogonal complement $\tilde{\beta}$ in the equation above, we find that
\begin{equation}
    \Tr \rho \alpha \tilde{\beta} \alpha = \Tr \Lambda_\alpha(\rho) \tilde{\beta} \alpha  \tilde{\beta} + \Delta {\cal F} [ \Lambda_\alpha(\rho), \tilde{\beta},\alpha].
    \label{eq:case2}
\end{equation}
On comparing Eqs. (\ref{eq:case1}) and (\ref{eq:case2}), we find that
\begin{equation}
    \Delta {\cal F} [ \Lambda_\alpha(\rho), \tilde{\beta},\alpha] = \Delta {\cal F} [ \Lambda_\alpha(\rho), \beta,\alpha].
    \label{eq:specialsymmetry}
\end{equation}
Thus, we see that the change of the quasi-probability is invariant under the exchange of the projector that is measured first (the one causing ``disturbance'' to the other) and its orthogonal complement.

We now assume  that the symmetry (\ref{eq:specialsymmetry}) applies to any state $\rho$. In reference to an opposite measurement order of that in Eq. (\ref{eq:specialsymmetry}), the third condition on the quasi-probability is:
\paragraph{Condition 3.}
\begin{equation}
    \Delta {\cal F} (\rho, \tilde{\alpha},\beta) = \Delta {\cal F}(\rho,\alpha,\beta).
    \label{eq:disturbancesymmetry}
\end{equation}

Next, we calculate the marginal of the change $\Delta {\cal F}$ over $\alpha$. By using Eqs. (\ref{eq:correctmarginals2}) and (\ref{eq:realquasi}) we find that
\begin{eqnarray}
   \Delta {\cal F}(\rho,\alpha,\beta) + \Delta {\cal F}(\rho,\tilde{\alpha},\beta) = \frac{1}{2} \left [ \Tr \rho \beta - \Tr \Lambda_\alpha(\rho) \beta \right ].
    \label{eq:deltamarginal}
\end{eqnarray}
By inserting (\ref{eq:disturbancesymmetry}) into (\ref{eq:deltamarginal}) we find that
\begin{equation}
    \Delta {\cal F}(\rho,\alpha,\beta) = \frac{1}{2} \left [ \Tr \rho \beta - \Tr \Lambda_\alpha(\rho) \beta \right ].
    \label{eq:mhchange}
\end{equation}
Therefore we have \cite{Johansen-Quantheosuccproj:07}
\begin{equation}
    {\cal F}(\rho,\alpha,\beta) = \Tr \rho \alpha \beta \alpha +
    \frac{1}{2} \left [ \Tr \rho \beta - \Tr \Lambda_\alpha(\rho) \beta \right ].
    \label{eq:mh}
\end{equation}
This is the resulting pre-measurement quasi-probability. We see that it differs from the post-measurement joint probability only if the probability for $\beta$ is disturbed by the preceding measurement of $\alpha$. The only possibility for the pre-measurement quasi-probability to become negative is that the measurement disturbance is sufficiently large.

By inserting (\ref{eq:lueders}), Eq. (\ref{eq:mh}) may also be written in the form
\begin{equation}
    {\cal F}(\rho,\alpha,\beta) = \frac{1}{2} \Tr \left [ \rho \left ( \alpha \beta + \beta \alpha \right ) \right ].
    \label{eq:mhsymm}
\end{equation}
Here we can see that the quasi-probability indeed satisfies the condition (\ref{eq:ordersymmetry}) of order symmetry. So, order symmetry (\ref{eq:ordersymmetry}) follows from the assumption of disturbance symmetry (\ref{eq:disturbancesymmetry}).  This quasi-probability has been discussed by various authors \cite{Terletsky-claslimiquanmech:37,Margenau+Hill-CorrbetwMeasQuan:61}.

It follows from the analysis in Ref. \cite{Hartle-Lineposivirtprob:04} that the distribution (\ref{eq:mhsymm}) is bounded between $-1/8$ and $1$. The lower bound is reached when the preparation $\rho$ and the two projectors $\alpha$ and $\beta$ correspond to socalled trine states \cite{Peres+Wootters-OptiDeteQuanInfo:91}.

One of the main purposes with a quasi-probability distribution is to provide an alternative representation of quantum states. It has been shown that the distribution (\ref{eq:mhsymm}) defined over classical phase space determines the state uniquely \cite{Sala+PalaoETAL-Phasspacformquan:97}. However, the informational content of this distribution is not complete in general. For example, in two-dimensional Hilbert space, the most general density matrix contains three real parameters. It may be shown that at least for a large class of observables, the distribution (\ref{eq:mhsymm}) contains fewer parameters.

In order to complete the information contained in (\ref{eq:mhsymm}), we may consider a complex extension of the distribution. A possible complex extension of (\ref{eq:mhsymm}) is rather obvious, since it is the real part of the complex expression $\Tr \rho \alpha \beta$. This is in fact a quasi-probability in its own right, giving correct marginal probabilities. It was first explored in phase space by Kirkwood \cite{Kirkwood-QuanStatAlmoClas:33}, and generalized to arbitrary observables by Dirac \cite{Dirac-AnalBetwClasQuan:45}. However, Dirac could not find the physical interpretation of this expression, and so he did not pursue it further.

Here, we shall derive the complex extension of (\ref{eq:mhsymm}) by some further conditions. Thus, we seek a complex quasi-probability
\begin{equation}
    {\cal G}(\rho,\alpha,\beta) = {\cal F}(\rho,\alpha,\beta) + i \, {\cal I} (\rho,\alpha,\beta).
\end{equation}
where the real part ${\cal F}(\rho,\alpha,\beta)$ is given by (\ref{eq:mhsymm}) and the imaginary part ${\cal I}(\rho,\alpha,\beta)$ is to be determined.

The argument leading to Eq. (\ref{eq:postdistribution}) still stands. This implies

\paragraph{Condition 4.}
\begin{equation}
    {\cal I} [\Lambda_\alpha(\rho),\alpha,\beta] = 0.
    \label{eq:imaginaryvanish}
\end{equation}

The complex quasi-probability must fulfill the same marginality conditions as the real distribution. Otherwise, it would lead to complex expectation values of hermitian observables. We must therefore impose the following:

\paragraph{Condition 5.}
\begin{subequations}
\begin{eqnarray}
      {\cal I}(\rho,\alpha,\beta) + {\cal I}(\rho,\alpha,\tilde{\beta}) &=& 0,
      \label{eq:imaginarymarginals1}\\
      {\cal I}(\rho,\alpha,\beta) + {\cal I}(\rho,\tilde{\alpha},\beta) &=& 0.
      \label{eq:imaginarymarginals2}
\end{eqnarray}
    \label{eq:imaginarymarginals}
\end{subequations}

Classically, the pre-measurement state and the post-measurement state is the same. That is why joint probabilities may be defined directly in terms of successive measurements. Quantum mechanically, a number of different pre-measurement states will give rise to the same post-measurement state $\Lambda_\alpha(\rho)$. We will now first find an equivalence class of pre-measurement states that gives rise to the same post-measurement state. To this end, we introduce the unitary operator $e^{i \phi \alpha}$. Since $e^{i \phi \alpha} \gamma = \gamma$ if $\alpha$ and $\gamma$ are orthogonal projectors, and $e^{i \phi \alpha} \alpha = e^{i \phi} \alpha$,
this operator will be denoted as a selective phase rotation operator. A selective phase rotation of the initial state $\rho$ gives the state $\rho^\phi_\alpha = e^{i \phi \alpha} \rho e^{- i \phi \alpha}$. We notice that the post-measurement state (\ref{eq:lueders}) may be written in the form \cite{Johansen-Recoweakvaluwith:07} $\Lambda_\alpha(\rho) = \left ( \rho + \rho^\pi_\alpha \right )/2$. This is a state consisting of a classical mixture of the original state $\rho$ and the same state selectively phase rotated an angle $\pi$. Thus, the phase related to the projector $\alpha$ has been completely randomized. This is an example of complete decoherence, but only with respect to a single projector (note that the complement $\tilde{\alpha}$ has also decohered).

Since the L\"uders map $\Lambda_\alpha$ entails phase randomization due to the action of the selective phase rotation operator $e^{i \phi \alpha}$, a selective phase rotation of the pre-measurement state $\rho$ does not alter the post-measurement state $\Lambda_\alpha(\rho)$. Hence, it can be shown that
$\Lambda_\alpha(\rho^\phi_\alpha) = \Lambda_\alpha(\rho)$. Furthermore, any state $\rho^\phi_\alpha$ gives rise to the same joint probability in a successive measurement of $\alpha$ and $\beta$, $\Tr \rho^\phi_\alpha \alpha \beta \alpha = \Tr \rho \alpha \beta \alpha$. Therefore, it is not possible to distinguish the states $\rho^\phi_\alpha$ by measuring $\alpha$ and $\beta$ successively. We shall refer to them as ``classically equivalent'' with respect to the successive measurement of $\alpha$ and $\beta$. They also give rise to the same post-measurement distribution ${\cal F} \left [ \Lambda_\alpha(\rho^\phi_\alpha),\alpha,\beta \right ] = \Tr \rho \alpha \beta \alpha$. However, they give rise to different changes of the quasi-probability $\Delta {\cal F} (\rho^\phi_\alpha,\alpha,\beta)$.

We have already seen that the imaginary term ${\cal I}(\rho,\alpha,\beta)$ should be a `` disturbance term'' in the sense of Eq. (\ref{eq:imaginaryvanish}). The simplest possibility is to assume that it takes the same form as the real disturbance term $\Delta {\cal F} (\sigma,\alpha,\beta)$ for some state $\sigma$. This would imply that Eqs. (\ref{eq:imaginaryvanish}) and (\ref{eq:imaginarymarginals1}) are automatically satisfied. However, we must have $\sigma \ne \rho$. Otherwise, there would be a conflict between conditions (\ref{eq:deltamarginal}) and (\ref{eq:imaginarymarginals2}). Our assumption will be that ${\cal I}(\rho,\alpha,\beta)$ should take the same form as the real disturbance term (\ref{eq:mhchange}), but for one of the classically equivalent states $\rho^\phi_\alpha$,

\paragraph{Condition 6.}
\begin{equation}
    {\cal I}(\rho,\alpha,\beta) = \frac{1}{2} \left [ \Tr \rho^\phi_\alpha \beta - \Tr \Lambda_\alpha(\rho) \beta \right ].
    \label{eq:imaginarydisturbance}
\end{equation}
By combining Eqs. (\ref{eq:imaginarymarginals2}) and (\ref{eq:imaginarydisturbance}), we find that \mbox{$\phi=\pi/2$}. Therefore, the complex quasi-probability may be written in the form \cite{Johansen-Quantheosuccproj:07}
\begin{eqnarray}
    {\cal G}(\rho,\alpha,\beta) = \Tr \rho \alpha \beta &+& \frac{1}{2} \left [ \Tr \rho \alpha \beta - \Tr \Lambda_\alpha(\rho) \beta \right ] \nonumber \\ &+& \frac{i}{2} \left [ \Tr \rho^{\frac{\pi}{2}}_\alpha \beta - \Tr \Lambda_\alpha(\rho) \beta \right ].
    \label{eq:complexcomplex}
\end{eqnarray}
On inserting the expression (\ref{eq:lueders}) for $\Lambda_\alpha(\rho)$ we find that Eq. (\ref{eq:complexcomplex}) simplifies to
\begin{equation}
    {\cal G}(\rho,\alpha,\beta) = \Tr \rho \alpha \beta.
    \label{eq:dirac}
\end{equation}
This is recognized as the complex quasi-probability discussed by Dirac \cite{Dirac-AnalBetwClasQuan:45}. Note that the sign in Eq. (\ref{eq:imaginarydisturbance}) was arbitrarily chosen. If an opposite sign had been assigned, we would have found the quasi-probability $\Tr \rho \beta \alpha $. We have found no way of distinguishing between these two alternatives. We also note that (\ref{eq:dirac}) does not fulfill the order symmetry (\ref{eq:ordersymmetry}), but the condition
\begin{equation}
    {\cal G}(\rho,\beta,\alpha) = {\cal G}(\rho,\alpha,\beta)^*.
\end{equation}
Exchanging the measurement order is equivalent to complex conjugation.

The distribution (\ref{eq:dirac}) is informationally complete, i.e., it determines the density matrix uniquely, provided that the two projectors belongs to two different projection valued measures where every projector is trace 1 and the two projection valued measures have no common elements \cite{Johansen-Quantheosuccproj:07}. This means that any pair of nondegenerate observables describe a quantum state completely provided only that they cannot both have a well-defined value for any state whatsoever. This is a beautiful illustration of the principle of complementarity.

The quasi-probability (\ref{eq:dirac}) is closely related to weak measurements \cite{Steinberg-Condprobquantheo:95}.  The real part may be observed directly in terms the correlations between pointer positions in successive measurements of the two projectors, provided that the interaction with the first pointer is sufficiently weak \cite{Johansen+Mello}. This scheme requires no post-selection. This is a different manifestation of the uniqueness of this distribution.

The distribution (\ref{eq:dirac}) may find applications e.g. in the theory of quantum information, communication and computing, where the goal is to construct procedures that outperform any classical counterpart. Due to its informational completeness, it may become a useful tool in information retrieval from successive measurements. It can also be mentioned that (\ref{eq:dirac}) has been discussed in connection with linearly positive histories \cite{Goldstein+Page-LinePosiHist:95,Diosi-AnomofWeDecoCrit:04,
Hartle-Lineposivirtprob:04,Marlow-Bayeaccoquanhist:06}.

The derivation of the real quasi-probability (\ref{eq:mhsymm}) relies on the assumption (\ref{eq:disturbancesymmetry}) of disturbance symmetry. The distribution (\ref{eq:mhsymm}) satisfies order symmetry (\ref{eq:ordersymmetry}). Thus, order symmetry (\ref{eq:ordersymmetry}) follows from disturbance symmetry (\ref{eq:disturbancesymmetry}). We demonstrated that for particular class of states, disturbance symmetry (\ref{eq:disturbancesymmetry}) could be derived from order symmetry (\ref{eq:ordersymmetry}). It is still an open question whether disturbance symmetry (\ref{eq:disturbancesymmetry}) may be derived from order symmetry (\ref{eq:ordersymmetry}) for arbitrary states.

In conclusion, we have derived a unique quasi-probability for arbitrary projectors from the analysis of successive projective measurements. Nonclassical properties of the quasi-probability is closely related to measurement disturbance. It unites the concept of projective and weak measurements in a common formalism.

The author acknowledges many stimulating talks with Pier A. Mello. Preliminary versions of this work have been presented elsewhere \cite{Johansen-Weakvaluprojmeas:07,Johansen-Negaprobmeasdist:08}.


\begin{thebibliography}{27}
\expandafter\ifx\csname natexlab\endcsname\relax\def\natexlab#1{#1}\fi
\expandafter\ifx\csname bibnamefont\endcsname\relax
  \def\bibnamefont#1{#1}\fi
\expandafter\ifx\csname bibfnamefont\endcsname\relax
  \def\bibfnamefont#1{#1}\fi
\expandafter\ifx\csname citenamefont\endcsname\relax
  \def\citenamefont#1{#1}\fi
\expandafter\ifx\csname url\endcsname\relax
  \def\url#1{\texttt{#1}}\fi
\expandafter\ifx\csname urlprefix\endcsname\relax\def\urlprefix{URL }\fi
\providecommand{\bibinfo}[2]{#2}
\providecommand{\eprint}[2][]{\url{#2}}

\bibitem[{\citenamefont{Bell}(1964)}]{Bell-EinsPodoRosepara:64}
\bibinfo{author}{\bibfnamefont{J.~S.} \bibnamefont{Bell}},
  \bibinfo{journal}{Physics (Long Island City, N.Y.)}
  \textbf{\bibinfo{volume}{1}}, \bibinfo{pages}{195} (\bibinfo{year}{1964}).

\bibitem[{\citenamefont{Wigner}(1971)}]{Wigner-QuanMechDistFunc:71}
\bibinfo{author}{\bibfnamefont{E.~P.} \bibnamefont{Wigner}}, in
  \emph{\bibinfo{booktitle}{Perspectives in Quantum Theory}}, edited by
  \bibinfo{editor}{\bibfnamefont{W.}~\bibnamefont{Yourgrau}} \bibnamefont{and}
  \bibinfo{editor}{\bibfnamefont{A.}~\bibnamefont{{van der Merwe}}}
  (\bibinfo{publisher}{MIT Press}, \bibinfo{address}{Cambridge},
  \bibinfo{year}{1971}), pp. \bibinfo{pages}{25--36}.

\bibitem[{\citenamefont{Schwinger}(2001)}]{Schwinger-QuanMech:01a}
\bibinfo{author}{\bibfnamefont{J.}~\bibnamefont{Schwinger}},
  \emph{\bibinfo{title}{Quantum Mechanics: Symbolism of Atomic Measurements}}
  (\bibinfo{publisher}{Springer}, \bibinfo{year}{2001}), pp.
  \bibinfo{pages}{10--14}.

\bibitem[{\citenamefont{Wigner}(1932)}]{Wigner-QuanCorrTherEqui:32}
\bibinfo{author}{\bibfnamefont{E.}~\bibnamefont{Wigner}},
  \bibinfo{journal}{Phys. Rev.} \textbf{\bibinfo{volume}{40}},
  \bibinfo{pages}{749} (\bibinfo{year}{1932}).

\bibitem[{\citenamefont{Gibbons et~al.}(2004)\citenamefont{Gibbons, Hoffman,
  and Wootters}}]{Gibbons+Wootters-Discphasspacbase:04}
\bibinfo{author}{\bibfnamefont{K.~S.} \bibnamefont{Gibbons}},
  \bibinfo{author}{\bibfnamefont{M.~J.} \bibnamefont{Hoffman}},
  \bibnamefont{and} \bibinfo{author}{\bibfnamefont{W.~K.}
  \bibnamefont{Wootters}}, \bibinfo{journal}{Phys. Rev. A}
  \textbf{\bibinfo{volume}{70}}, \bibinfo{pages}{062101}
  (\bibinfo{year}{2004}).

\bibitem[{\citenamefont{Dirac}(1945)}]{Dirac-AnalBetwClasQuan:45}
\bibinfo{author}{\bibfnamefont{P.~A.~M.} \bibnamefont{Dirac}},
  \bibinfo{journal}{Rev. Mod. Phys.} \textbf{\bibinfo{volume}{17}},
  \bibinfo{pages}{195} (\bibinfo{year}{1945}).

\bibitem[{\citenamefont{Feynman}(1982)}]{Feynman-SimuPhyswithComp:82}
\bibinfo{author}{\bibfnamefont{R.~P.} \bibnamefont{Feynman}},
  \bibinfo{journal}{Int. J. Theor. Phys.} \textbf{\bibinfo{volume}{21}},
  \bibinfo{pages}{467} (\bibinfo{year}{1982}).

\bibitem[{\citenamefont{Leonhardt}(1997)}]{Leonhardt-MeasQuanStatLigh:97}
\bibinfo{author}{\bibfnamefont{U.}~\bibnamefont{Leonhardt}},
  \emph{\bibinfo{title}{Measuring the Quantum State of Light}}
  (\bibinfo{publisher}{Cambridge University Press},
  \bibinfo{address}{Cambridge}, \bibinfo{year}{1997}).

\bibitem[{\citenamefont{Vogel et~al.}(2001)\citenamefont{Vogel, Welsch, and
  Wallentowitz}}]{Vogel+WelschETAL-QuanOpti:01}
\bibinfo{author}{\bibfnamefont{W.}~\bibnamefont{Vogel}},
  \bibinfo{author}{\bibfnamefont{D.-G.} \bibnamefont{Welsch}},
  \bibnamefont{and}
  \bibinfo{author}{\bibfnamefont{S.}~\bibnamefont{Wallentowitz}},
  \emph{\bibinfo{title}{Quantum Optics: An Introduction}}
  (\bibinfo{publisher}{Wiley-VCH}, \bibinfo{address}{Berlin},
  \bibinfo{year}{2001}), \bibinfo{edition}{2nd} ed.

\bibitem[{\citenamefont{Aharonov et~al.}(1988)\citenamefont{Aharonov, Albert,
  and Vaidman}}]{Aharonov+AlbertETAL-ResuMeasCompSpin:88}
\bibinfo{author}{\bibfnamefont{Y.}~\bibnamefont{Aharonov}},
  \bibinfo{author}{\bibfnamefont{D.~Z.} \bibnamefont{Albert}},
  \bibnamefont{and} \bibinfo{author}{\bibfnamefont{L.}~\bibnamefont{Vaidman}},
  \bibinfo{journal}{Phys. Rev. Lett.} \textbf{\bibinfo{volume}{60}},
  \bibinfo{pages}{1351} (\bibinfo{year}{1988}).

\bibitem[{\citenamefont{L\"{u}ders}(1951)}]{Lueders-UEbeZust:51}
\bibinfo{author}{\bibfnamefont{G.}~\bibnamefont{L\"{u}ders}},
  \bibinfo{journal}{Ann. Physik} \textbf{\bibinfo{volume}{8}},
  \bibinfo{pages}{322} (\bibinfo{year}{1951}).

\bibitem[{\citenamefont{Wigner}(1963)}]{Wigner-ProbMeas:63}
\bibinfo{author}{\bibfnamefont{E.~P.} \bibnamefont{Wigner}},
  \bibinfo{journal}{Am. J. Phys.} \textbf{\bibinfo{volume}{31}},
  \bibinfo{pages}{6} (\bibinfo{year}{1963}).

\bibitem[{\citenamefont{Johansen}(2007{\natexlab{a}})}]{Johansen-Quantheosuccp%
roj:07}
\bibinfo{author}{\bibfnamefont{L.~M.} \bibnamefont{Johansen}},
  \bibinfo{journal}{Phys. Rev. A} \textbf{\bibinfo{volume}{76}},
  \bibinfo{pages}{012119} (\bibinfo{year}{2007}{\natexlab{a}}).

\bibitem[{\citenamefont{Terletsky}(1937)}]{Terletsky-claslimiquanmech:37}
\bibinfo{author}{\bibfnamefont{Y.~P.} \bibnamefont{Terletsky}},
  \bibinfo{journal}{Zh. Eksp. Teor. Fiz.} \textbf{\bibinfo{volume}{7}},
  \bibinfo{pages}{1290} (\bibinfo{year}{1937}).

\bibitem[{\citenamefont{Margenau and
  Hill}(1961)}]{Margenau+Hill-CorrbetwMeasQuan:61}
\bibinfo{author}{\bibfnamefont{H.}~\bibnamefont{Margenau}} \bibnamefont{and}
  \bibinfo{author}{\bibfnamefont{R.~N.} \bibnamefont{Hill}},
  \bibinfo{journal}{Prog. Theor. Phys.} \textbf{\bibinfo{volume}{26}},
  \bibinfo{pages}{722} (\bibinfo{year}{1961}).

\bibitem[{\citenamefont{Hartle}(2004)}]{Hartle-Lineposivirtprob:04}
\bibinfo{author}{\bibfnamefont{J.~B.} \bibnamefont{Hartle}},
  \bibinfo{journal}{Phys. Rev. A} \textbf{\bibinfo{volume}{70}},
  \bibinfo{pages}{022104} (\bibinfo{year}{2004}).

\bibitem[{\citenamefont{Peres and
  Wootters}(1991)}]{Peres+Wootters-OptiDeteQuanInfo:91}
\bibinfo{author}{\bibfnamefont{A.}~\bibnamefont{Peres}} \bibnamefont{and}
  \bibinfo{author}{\bibfnamefont{W.~K.} \bibnamefont{Wootters}},
  \bibinfo{journal}{Phys. Rev. Lett.} \textbf{\bibinfo{volume}{66}},
  \bibinfo{pages}{1119} (\bibinfo{year}{1991}).

\bibitem[{\citenamefont{Sala et~al.}(1997)\citenamefont{Sala, Palao, and
  Muga}}]{Sala+PalaoETAL-Phasspacformquan:97}
\bibinfo{author}{\bibfnamefont{R.}~\bibnamefont{Sala}},
  \bibinfo{author}{\bibfnamefont{J.~P.} \bibnamefont{Palao}}, \bibnamefont{and}
  \bibinfo{author}{\bibfnamefont{J.}~\bibnamefont{Muga}},
  \bibinfo{journal}{Phys. Lett. A} \textbf{\bibinfo{volume}{231}},
  \bibinfo{pages}{304} (\bibinfo{year}{1997}).

\bibitem[{\citenamefont{Kirkwood}(1933)}]{Kirkwood-QuanStatAlmoClas:33}
\bibinfo{author}{\bibfnamefont{J.~G.} \bibnamefont{Kirkwood}},
  \bibinfo{journal}{Phys. Rev.} \textbf{\bibinfo{volume}{44}},
  \bibinfo{pages}{31} (\bibinfo{year}{1933}).

\bibitem[{\citenamefont{Johansen}(2007{\natexlab{b}})}]{Johansen-Recoweakvaluw%
ith:07}
\bibinfo{author}{\bibfnamefont{L.~M.} \bibnamefont{Johansen}},
  \bibinfo{journal}{Phys. Lett. A} \textbf{\bibinfo{volume}{366}},
  \bibinfo{pages}{374} (\bibinfo{year}{2007}{\natexlab{b}}).

\bibitem[{\citenamefont{Steinberg}(1995)}]{Steinberg-Condprobquantheo:95}
\bibinfo{author}{\bibfnamefont{A.~M.} \bibnamefont{Steinberg}},
  \bibinfo{journal}{Phys. Rev. A} \textbf{\bibinfo{volume}{52}},
  \bibinfo{pages}{32} (\bibinfo{year}{1995}).

\bibitem[{\citenamefont{Johansen and Mello}()}]{Johansen+Mello}
\bibinfo{author}{\bibfnamefont{L.~M.} \bibnamefont{Johansen}} \bibnamefont{and}
  \bibinfo{author}{\bibfnamefont{P.~A.} \bibnamefont{Mello}},
  \bibinfo{howpublished}{(unpublished)}.

\bibitem[{\citenamefont{Goldstein and
  Page}(1995)}]{Goldstein+Page-LinePosiHist:95}
\bibinfo{author}{\bibfnamefont{S.}~\bibnamefont{Goldstein}} \bibnamefont{and}
  \bibinfo{author}{\bibfnamefont{D.~N.} \bibnamefont{Page}},
  \bibinfo{journal}{Phys. Rev. Lett.} \textbf{\bibinfo{volume}{74}},
  \bibinfo{pages}{3715} (\bibinfo{year}{1995}).

\bibitem[{\citenamefont{Di\'{o}si}(2004)}]{Diosi-AnomofWeDecoCrit:04}
\bibinfo{author}{\bibfnamefont{L.}~\bibnamefont{Di\'{o}si}},
  \bibinfo{journal}{Phys. Rev. Lett.} \textbf{\bibinfo{volume}{92}},
  \bibinfo{pages}{170401} (\bibinfo{year}{2004}).

\bibitem[{\citenamefont{Marlow}(2006)}]{Marlow-Bayeaccoquanhist:06}
\bibinfo{author}{\bibfnamefont{T.}~\bibnamefont{Marlow}},
  \bibinfo{journal}{Ann. Phys.} \textbf{\bibinfo{volume}{321}},
  \bibinfo{pages}{1103} (\bibinfo{year}{2006}).

\bibitem[{\citenamefont{Johansen}(2007{\natexlab{c}})}]{Johansen-Weakvaluprojm%
eas:07}
\bibinfo{author}{\bibfnamefont{L.~M.} \bibnamefont{Johansen}}, contribution to
  \emph{\bibinfo{booktitle}{Weak Values and Weak Measurements: A New Approach
  to Reality in Quantum Theory}}, edited by
  \bibinfo{editor}{\bibfnamefont{P.}~\bibnamefont{Davies}},
  \bibinfo{organization}{Beyond Center, University of Arizona}
  (\bibinfo{publisher}{unpublished}, \bibinfo{year}{2007}{\natexlab{c}}).

\bibitem[{\citenamefont{Johansen}(2008)}]{Johansen-Negaprobmeasdist:08}
\bibinfo{author}{\bibfnamefont{L.~M.} \bibnamefont{Johansen}}, contribution to
  \emph{\bibinfo{booktitle}{APS March Meeting}}, \bibinfo{organization}{New
  Orleans} (\bibinfo{publisher}{unpublished}, \bibinfo{year}{2008}).

\end{thebibliography}
\end{document}